\documentclass[draft]{agujournal2019} 
\usepackage{url} 
\usepackage{lineno}
\usepackage[inline]{trackchanges} 
\usepackage{soul}
\journalname{Geophysical Research Letters}

\begin{document}

	\title{A New Mechanism for ULF Wave Modulation of Energetic Electron Precipitation}
	
	\authors{L.~Olifer\affil{1}, D.~Zhou\affil{1}, M.~Patel\affil{2}, I.R.~Mann\affil{1}, M.K.~Hudson\affil{2}, A.W.~Degeling\affil{3}, C.O.~Heinke\affil{1}, G.R.~Sivakoff\affil{1}, A.~Kale\affil{1}, S.~Kasahara\affil{4}, S.~Yokota\affil{5}, K.~Keika\affil{4}, T.~Hori\affil{6}, T.~Mitani\affil{7}, T.~Takashima\affil{7}, Y.~Kasahara\affil{8}, S.~Matsuda\affil{8}, A.~Shinbori\affil{6}, A.~Matsuoka\affil{9}, M.~Teramoto\affil{10}, K.~Yamamoto\affil{6}, I.~Shinohara\affil{7}, Y.~Miyoshi\affil{6}}
	
	\affiliation{1}{Department of Physics, University of Alberta, Edmonton, AB, Canada}
    \affiliation{2}{Department of Physics and Astronomy, Dartmouth College, Hanover, NH, USA}
    \affiliation{3}{Institute of Space Science, Shandong University, Jinan, China}
    \affiliation{4}{The University of Tokyo, Japan}
    \affiliation{5}{Osaka University, Japan}
    \affiliation{6}{Nagoya University, Japan}
    \affiliation{7}{Japan Aerospace Exploration Agency (JAXA), Japan}
    \affiliation{8}{Kanazawa University, Japan}
    \affiliation{9}{Kyoto University, Japan}
    \affiliation{10}{Kyushu Institute of Technology, Japan}

	\correspondingauthor{Leonid Olifer}{olifer@ualberta.ca}
	
	\begin{keypoints}
		\item Balloon-borne X-ray and ground-based riometer observations reveal $\sim$4 min period precipitation bursts phase-aligned with Pc5 ULF waves
		\item Arase satellite data show ~4 min energy-dispersed flux and chorus wave power modulations traced to the balloon location via drift dispersion
		\item ULF wave-driven flux modulations cross the Kennel-Petschek limit, triggering periodic chorus wave growth and periodic electron precipitation
	\end{keypoints}

	\begin{abstract}
		The May 2024 geomagnetic superstorm provided the opportunity to explore how strong wave-particle interactions affect energetic electron precipitation under intense driving. Using coordinated measurements from a balloon-borne Timepix-based X-ray detector, ground-based riometers and magnetometers, and Arase satellite observations, we identified quasi-periodic bursts of energetic electron precipitation coincident with Pc5 ultra low frequency (ULF) wave oscillations. Arase satellite data revealed energy-dispersed trapped energetic electron flux modulations in the `seed' energy range, indicating that trapped electron flux was likely modulated by ULF waves. This letter reveals that these flux enhancements surpassed the Kennel-Petschek (K-P) limit, creating intense chorus waves and driving periodic electron precipitation. Drift-dispersion analysis traced these modulations back to a source in the post-noon magnetospheric sector, matching balloon and ground-based measurements. Here, we propose a novel indirect ULF wave-driven mechanism for modulated energetic electron precipitation, whereby periodic modulations of `seed' electron fluxes enhance electron losses. 
	\end{abstract}

    \section*{Plain Language Summary}
        The May 2024 geomagnetic superstorm gave us a unique opportunity to test theories about how extreme space weather shapes the near-Earth environment. During such storms, waves in Earth's magnetic field can interact with energetic electrons in space, sometimes driving them into the atmosphere, where they affect both natural weather patterns as well as technology. In this study, we combined data from a high-altitude balloon, ground-based sensors, and the Arase satellite to investigate why there were bursts of electrons raining into the atmosphere every four minutes during the storm. We focused on the role of ultra-low-frequency (ULF) magnetic waves and how they might have created this unusual pattern. We discovered that ULF waves did not directly scatter electrons into the atmosphere. Instead, they periodically boosted the population of trapped electrons beyond a natural stability limit. This triggered the creation of other electromagnetic plasma waves known as chorus, which then drove electrons downward in regular bursts. Our findings reveal a new indirect mechanism for electron precipitation during storms, improving our understanding of how space weather operates. This knowledge is essential for predicting and mitigating the impacts of extreme storms on satellites, communications, and other critical technologies.

    \section{Introduction}
        The May 2024 geomagnetic superstorm was an intense and globally impactful space weather event driven by multiple successive coronal mass ejections from solar active region AR13664 \cite<e.g.,>[]{Spogli_2024}. It reached a minimum Dst of $-$412~nT, the lowest value since 1989, with elevated geomagnetic activity persisting for several days \cite<e.g.,>[and references therein]{Lawrence_2025, Hayakawa_2025}. The geomagnetic storm led to severe magnetopause compression below L-shell of 6~$R_E$ \cite{Gomez_2025}, very low-latitude auroral displays \cite{Grandin_2024}, significant ionospheric disruptions \cite{Singh_2024}, and potentially substantially increased orbital drag affecting the longevity of low Earth orbit satellites \cite{Parker_2024, Ashruf_2025}. Interestingly, damage to electric power distribution infrastructure appears to have remained limited across the planet during this geomagnetic storm. For example, \citeA{Lawrence_2025} and \cite{Caraballo_2025} reported a peak geomagnetically induced currents (GIC) remaining below 50~A in mid- and low-latitude regions. Characterizing storms of this magnitude is important both for advancing our understanding of extreme space weather and for designing appropriate mitigation strategies against the technological impacts arising from such events. In this paper, we examine a period of energetic particle dynamics and related bursts of energetic electron precipitation that occurred during the May 2024 storm. 


        Advance geomagnetic storm warnings were issued on 9~May~2024 by the NOAA space-weather prediction center (SWPC), initially for a potential G4 event, providing roughly one day of lead time for instigating a campaign of coordinated observations. Capitalizing on this window, \citeA{Olifer_2025} launched a novel high altitude balloon-borne Timepix-based detector during the storm's recovery phase, successfully recording sub-second resolution X-ray fluxes generated by galactic cosmic rays (GCRs) and by precipitating energetic electrons. Their balloon measurements, validated by concurrent riometer observations, identified clear quasi-periodic bursts of electron precipitation closely phased with monochromatic Pc5 ultra low frequency (ULF) oscillations observed both on the ground and in space. \citeA{Olifer_2025} hypothesized that these ULF waves played a major role in creating these periodic bursts of precipitation, but did not examine the event further. 
        
        In this paper, we distinguish between two different mechanisms for the ULF waves to cause modulated electron precipitation: direct and indirect. For example, \citeA{Rae_2018} and \citeA{Patel_2025} showed that ULF waves can change the size of the equatorial loss cone, thus directly causing modulated energetic particle precipitation. \citeA{Chaston_2018} showed that broadband electromagnetic waves, which include the ULF frequency range, may directly resonate with trapped energetic electrons and directly lead to particle precipitation \cite<see also>[]{Chaston_2015}. On the other hand, ULF waves have also been shown to modulate higher frequency plasma waves that themselves cause particle precipitation, thus indirectly modulating energetic electron scattering losses into the atmosphere. For example, ULF waves have been shown to modulate electromagnetic ion cyclotron waves \cite<EMIC, e.g.,>[]{Loto_aniu_2009}, whistler-mode chorus \cite<e.g.,>[and references therein]{Halford_2015, Watt_2011, Shang_2021}, and hiss waves \cite<e.g.,>[]{Breneman_2015}. The exact mechanisms responsible for the plasma wave modulation can vary depending on the event, but could include ULF wave-induced changes to the background conditions like magnetic field strength and cold plasma density \cite<e.g.,>[]{Spanswick_2005, Loto_aniu_2009}, or to the growth rates of the plasma waves through modulation of particle temperature anisotropy \cite<e.g.,>[]{Halford_2015}. 
                
        In this letter, we build upon these earlier findings by introducing a new mechanism of how ULF waves may indirectly modulate electron precipitation by periodically varying the flux levels of newly injected low-energy ($<$300 keV) electron populations. We show evidence for ULF wave modulation of background flux, which periodically exceeded the Kennel-Petschek (K-P) limit \cite<e.g.,>[]{Kennel_1966, Olifer_2021, Mourenas_2024}. This periodically changes the chorus wave growth rates and causes indirect ULF wave modulated precipitation.           

    \section{Data and Methodology}
        \subsection{The May 2024 Geomagnetic Superstorm}
\begin{figure}
                \centering
                \makebox[\textwidth][c]{\includegraphics[width=1.4\linewidth]{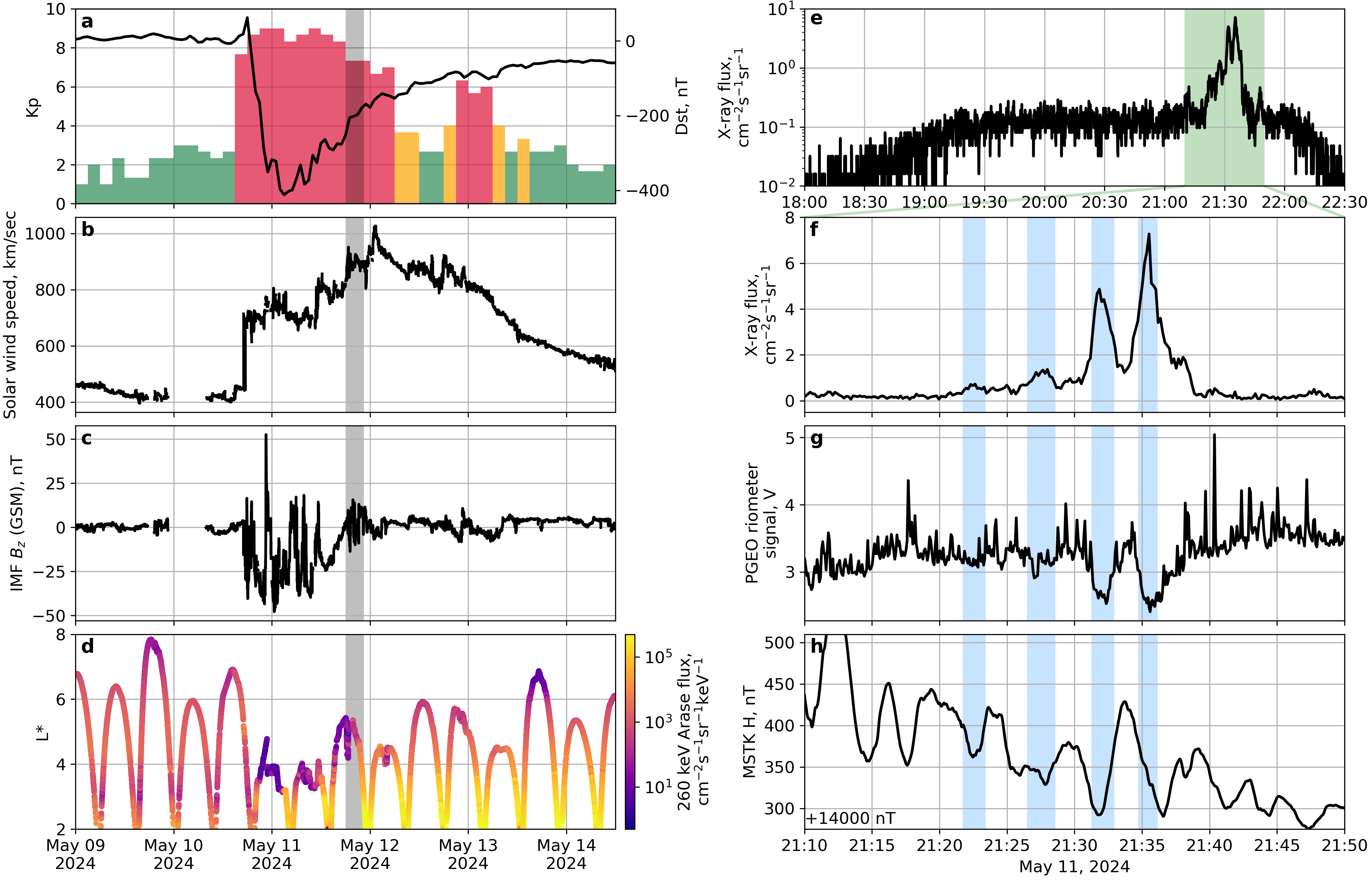}}
                \caption{Overview of the May 2024 superstorm and co-located observations of ULF-modulated electron precipitation. Left column shows the overview of the geomagnetic storm progression: (a) Kp index (bars) and Dst index (solid line, right axis), (b) OMNI solar‐wind speed, (c) $B_z$ component of the interplanetary magnetic field in GSM coordinates, and (d) Arase measurements of the 260~keV differential electron flux as a function of $L$*; the gray band denotes the interval of the balloon flight expanded in (e). (e) Atmospheric X-ray flux measured by a balloon-borne Timepix instrument; the green bar marks the interval containing identified modulated electron precipitation expanded in (f)-(g). (f) zoom showing four $\sim$4~min precipitation peaks (blue shading marks full-width at half-maximum); (g) Prince George (PGEO) riometer signal; and (h) Ministik Lake (MSTK) ground magnetometer H-component (offset by +14,000~nT), revealing magnetic Pc5 ULF wave oscillations that coincide with the precipitation peaks.}
                \label{fig:May_storm}
            \end{figure}

            Characterized by a Kp index of 9 (G5 on NOAA's geomagnetic storm scale) and a minimum Dst index of $-412$~nT (Figure~\ref{fig:May_storm}a), the May 2024 geomagnetic storm was the most intense G5 storm since March 1989. The left panel of Figure~\ref{fig:May_storm} summarizes selected geomagnetic indices and solar wind conditions during the May 2024 superstorm, as well as the 260~keV energetic electron dynamics as observed by the Arase satellite \cite{Miyoshi_2018}. The left column shows the storm progression from the onset to the end of the recovery phase. The main phase of the storm is characterized by the dropout of the 260~keV energetic electrons at around 18~UT on May~10, which is followed by two enhancement periods. The first enhancement occurred early in the day on May 11, increasing fluxes at $L$*$<$3. The second enhancement occred at $\sim$21~UT, increasing fluxes at L$*>$4 to levels approximately an order of magnitude above the pre-storm levels. \citeA{Olifer_2025} conducted the balloon flight during the second enhancement period. 

        \subsection{Balloon-borne Timepix X-ray Measurements}
            We examine a period of energetic electron precipitation using balloon-borne X-ray data collected during the \citeA{Olifer_2025} weather balloon launch campaign. The \citeA{Olifer_2025} payload used a first-generation Timepix-based silicon pixelated detector, and recorded bremsstrahlung X-rays produced by energetic electron precipitation during the balloon flight. Operating in Time-over-Threshold (ToT) mode with a 0.3~s exposure duration, the detector measured energy deposition in each 55~$\mu$m $\times$ 55~$\mu$m pixel across a 256 $\times$ 256 pixel matrix. The 500~$\mu$m-thick silicon sensor provided sensitivity primarily to photons in the 2-70 keV energy range, with the lower limit set by the electronic noise threshold and the upper limit constrained by detector efficiency. See \citeA{Olifer_2025} for more details on the launch campaign and the data processing. To limit the contamination of the X-ray signals from electrons and other light charged particles in this study, we also implemented an X-ray grading algorithm adapted from techniques commonly used for CCD imaging in X-ray astrophysics \cite<e.g.,>[]{Gendreau_1995}. X-ray grading uses the pattern of deposited X-ray charge in the detector to distinguish between real X-rays and charged particles. We used the ASCA (Advanced Satellite for Cosmology and Astrophysics) grading scheme \cite{Gendreau_1995}, selecting patterns corresponding to grades 0, 2, 3, 4, and 6 as likely X-ray detections, and grades 1, 5, and 7 as likely charged particle detections.  
            
            The right column of Figure~\ref{fig:May_storm} highlights the principal findings from \citeA{Olifer_2025}, focusing on a 40-minute interval during the storm's recovery phase when the balloon detected four distinct quasi-periodic bursts of energetic electron precipitation. These precipitation peaks exhibit a characteristic $\sim$4-minute periodicity that is closely matched by monochromatic Pc5 ULF wave oscillations observed by ground-based magnetometers \cite<cf. Figure~\ref{fig:May_storm}h and see>[for more details]{Olifer_2025}. Concurrent ground-based riometer measurements for Prince George (PGEO, Figure~\ref{fig:May_storm}g) station, representative of electron precipitation with energies $>$30~keV \cite{Spanswick_2013}, further confirmed the enhanced ionospheric absorption associated with these energetic electron precipitation events. Determining the physical cause for this periodic precipitation is the main goal of this paper.

        \subsection{Arase Satellite Data}
            In this study, we analyze measurements from the Arase satellite \cite{Miyoshi_2018} collected during its inbound pass between 18:00 and 22:30 UT on 11 May 2024. Specifically, we use data from two particle instruments: the Medium-Energy Particle Experiments -- Electron Analyzer \cite<MEP-e,>[]{Kasahara_2018} and the High-Energy Electron Experiments \cite<HEP,>[]{Mitani_2018}, as well as from the Plasma Wave Experiment \cite<PWE,>[]{Kasahara_2018} and the Magnetic Field Experiment \cite<MGF,>[]{Matsuoka_2018}. The combined MEP-e and HEP datasets reveal modulated electron fluxes in the energy range of $\sim$80~keV to $\sim$300~keV, and which also exhibit clear energy-dependent drift dispersion (see Section~3 for details). Here, we analyze timing differences between peaks in different energy channels to infer the spatial and temporal origin of this modulation through particle tracing, as described below. Figure~\ref{fig:May_storm}d presents an overview of the Arase electron flux measurements in the 260 keV channel over the course of the May 2024 storm.

        \subsection{MHD Particle Tracing}
            The energy-dependent drift dispersion observed in the Arase energetic electron flux data can be used for back-tracing particles to examine the location and origin of the flux modulation, i.e., the location in magnetic local time (MLT) and the UT time at which ULF waves could have adiabatically perturbed the electron flux. During the initial modulation period, this would be expected to create a coherent in-phase response in all energy channels. However, following this initial ULF wave interaction, the electrons would have been expected to drift disperse depending on their energy. In particular, we can estimate the perturbation onset time, $t_0$, and its corresponding MLT origin, $\phi_0$, defined as the point where flux modulations across all energy channels are in phase, using the relation:

            \begin{equation}
                \phi_{sat}=\dot \phi \left( E, \alpha \right) \left[ t_m\left( E \right)-t_0 \right] + \phi_0 \Rightarrow t_m=\frac{\phi_{sat} - \phi_0}{\dot \phi}+t_0,
                \label{eq:disp}
            \end{equation}
            
            \noindent where, $\phi_{sat}$ is the MLT of the satellite measurement, $t_m(E)$ is the measurement time of the modulation peak (or trough) at energy, $E$, and $\dot \phi$ is the electron's angular drift velocity, which depends on energy $E$ and pitch angle $\alpha$. Assuming a constant drift velocity along the dispersion path in the analysis presented here, $t_0$ and $\phi_0$ are obtained by performing a linear fit to the measured time-drift period, $\dot \phi^{-1}$, relation.

            We can further estimate the electron drift velocity using the simulation of the storm-time distorted magnetosphere for this storm from the Multiscale Atmosphere Geospace Environment (MAGE) whole geospace model simulation, which includes a global MHD model \cite<GAMERA,>[]{Zhang_2019, Sorathia_2020}, an inner magnetosphere model \cite<RCM,>[]{Toffoletto_2003}, and and ionospheric potential solver \cite<REMIX>[]{Merkin_2010}. MAGE is driven by upstream solar wind conditions from the WIND spacecraft. The resulting MHD fields are used to trace electron trajectories as test particles in the Dartmouth rbelt3d model \cite{Kress_2007}. One million test electrons were initiated on 11 May 2024 at 21:20 UT in the equatorial plane (a B-min surface) in the MHD fields between L-shells of 3 and 5, with energies between 100 and 500~keV, and with pitch angles between 5 and 175~deg. These electrons were traced using the guiding center approximation for 40~min. The average electron drift velocity was then estimated by calculating the time derivative of the electron location mapped along the field line to the equatorial plane.          

    \section{Results}
        \begin{figure}
            \centering
            \makebox[\textwidth][c]{\includegraphics[width=1\linewidth]{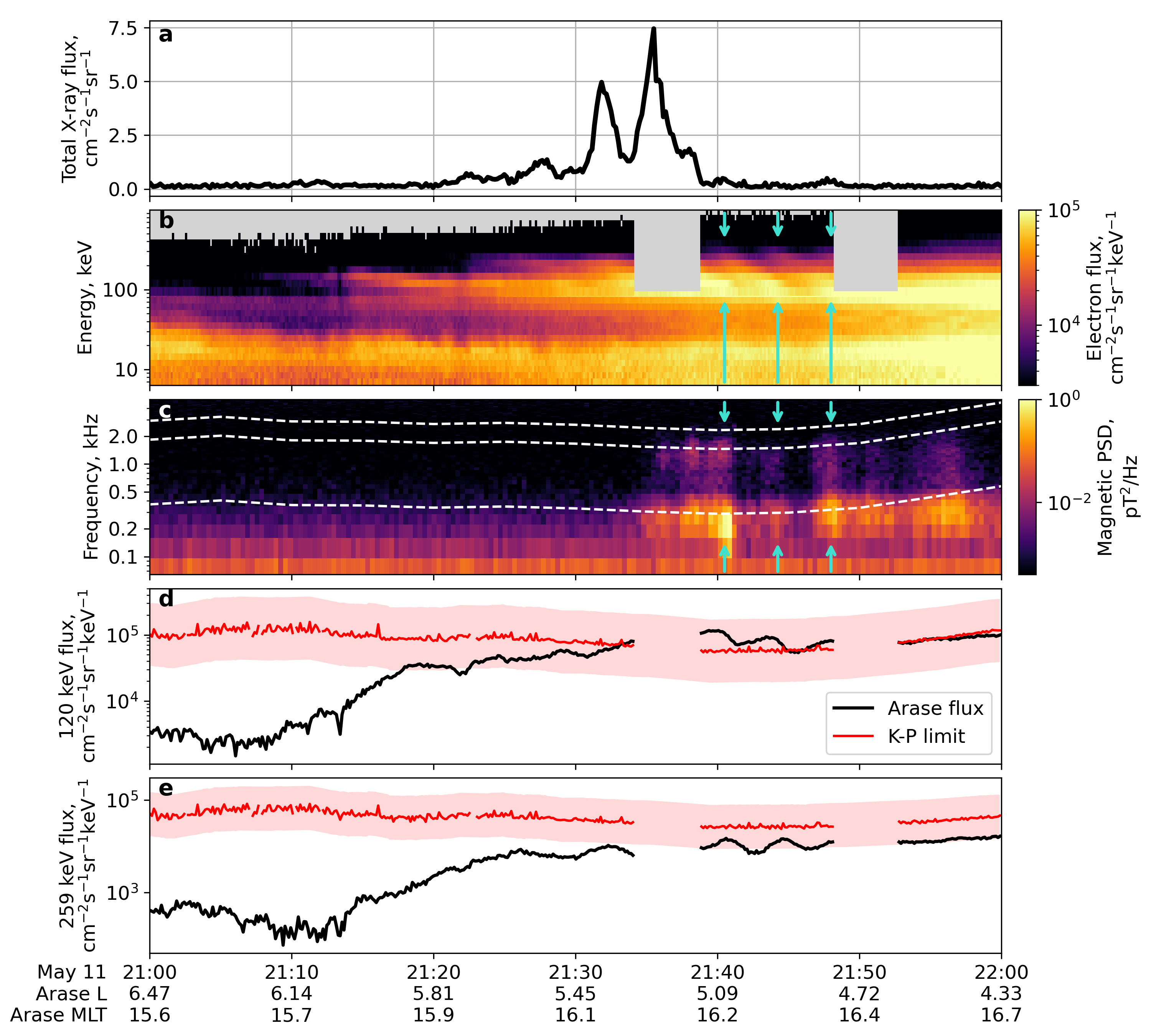}}
            \caption{Multi-instrument observations of energetic electron precipitation and trapped energetic electron flux modulation during the May 11, 2024, geomagnetic storm. (a) Total X-ray flux measured by the balloon-borne Timepix detector. (b) Energetic electron flux energy spectrum (combined MEP-e and HEP data) from the Arase satellite. (c) Magnetic power spectral density as measured by the Arase PWE instrument; dashed lines mark 0.1, 0.5, and 0.8 times the local electron gyrofrequency, $f_{ce}$, estimated from the magnetic field strength from the Arase MGF instrument. (d-e) Arase electron fluxes at 120 keV and 259 keV (black), respectively, overplotted with the theoretical relativistic Kennel-Petschek flux limit (red) derived from Arase data, including an assumed uncertainty range of a factor of 3 shown in pink shading. Vertical cyan lines mark the times of X-ray flux peaks observed by the balloon, shifted by 6 minutes to account for an estimated electron drift time from the balloon location to the Arase spacecraft.}
            \label{fig:Arase}
        \end{figure} 
    
        Figure~\ref{fig:Arase} shows a comparison between the balloon-borne X-ray observations (Figure~\ref{fig:Arase}a), the trapped energetic electron fluxes (Figure~\ref{fig:Arase}b, d, e), and VLF wave activity (Figure~\ref{fig:Arase}c) measured by the Arase satellite. Supplementary Figure~S1 shows the locations of the balloon and the Arase satellite during the event. The measured electron flux energy spectrum reveals energy-dependent dispersion within the newly injected electron population (Figure~\ref{fig:Arase}b, highlighted by vertical cyan lines). Higher-energy electrons arrive earlier in each modulation burst, as characterized in the dynamic energy spectrum by flux enhancements tracing trajectories from top left to bottom right in the spectrogram. Occurring at energies between $\sim$80-300 keV, such dispersion is consistent with evolution along a drift path after $\sim$10~min, corresponding to approximately 2–3~hr of drift in magnetic local time (MLT). The observed flux modulations also exhibit a characteristic repetition period of roughly 4~min. 
        
        Concurrent with these flux modulations, the magnetic plasma wave spectrum measured by the Arase satellite (Figure~\ref{fig:Arase}c) shows periodic $\sim$4-minute bursts of enhanced wave power concentrated near the typical chorus frequency band ($\sim$0.1-0.8 $f_{ce}$, where $f_{ce}$ is the local electron cyclotron frequency). Notably, Supplementary Figure~S2 shows that these enhanced VLF waves are resonant with $\sim$80-450~keV electrons with equatorial pitch angles below 75$^\circ$, i.e., the energetic electron population that exhibits energy dispersion described above. This suggests a compelling connection between the periodically enhanced flux of trapped energetic electrons, periodic chorus wave activity, and the periodic energetic electron precipitation observed by the balloon. It is possible that interaction with the ULF wave caused drift phase bunching of elelctrons, resulting in short periodic structures of the slowly drifting electrons. A drift resonance condition, $f=mf_d$ for a 4~min period ($f=$4.17~mHz) ULF wave and a characteristic 100~keV electron at L$=$5 ($\sim$91~min drift period) results in a relatively high $m=$23 mode number, corresponding to an azimuthal wavelength of 1.37~$R_E$. Interestingly, such waves are common and can be driven by drift-bounce resonance with ring current ions, which have been injected during the main phase \cite{Ozeke_2008}. 
        
        If the modulation of the electron flux is sufficiently large, it could create local regions where the flux is concentrated as a result of drift-phase bunching to such an extent that it locally exceeds the so-called Kennel-Petschek (K-P) limit. This theoretical upper flux limit proposed by \citeA{Kennel_1966} is established through an asymptotic balance between chorus wave growth and electron precipitation into the atmosphere. It also represents an approximate threshold between regimes of low and high chorus wave growth \cite{Chakraborty_2022}, as well as weak and strong pitch angle diffusion \cite{Olifer_2023, Ozeke_2024}. Indeed, Figures~\ref{fig:Arase}d and \ref{fig:Arase}e show that electron fluxes at 120~keV and 259~keV hover near or above the K-P limit obtained in relativistic formulation, implying that small flux modulations can trigger strong chorus wave amplification and hence strong precipitation.   
        
\begin{figure}
            \centering
            \makebox[\textwidth][c]{\includegraphics[width=1.2\linewidth]{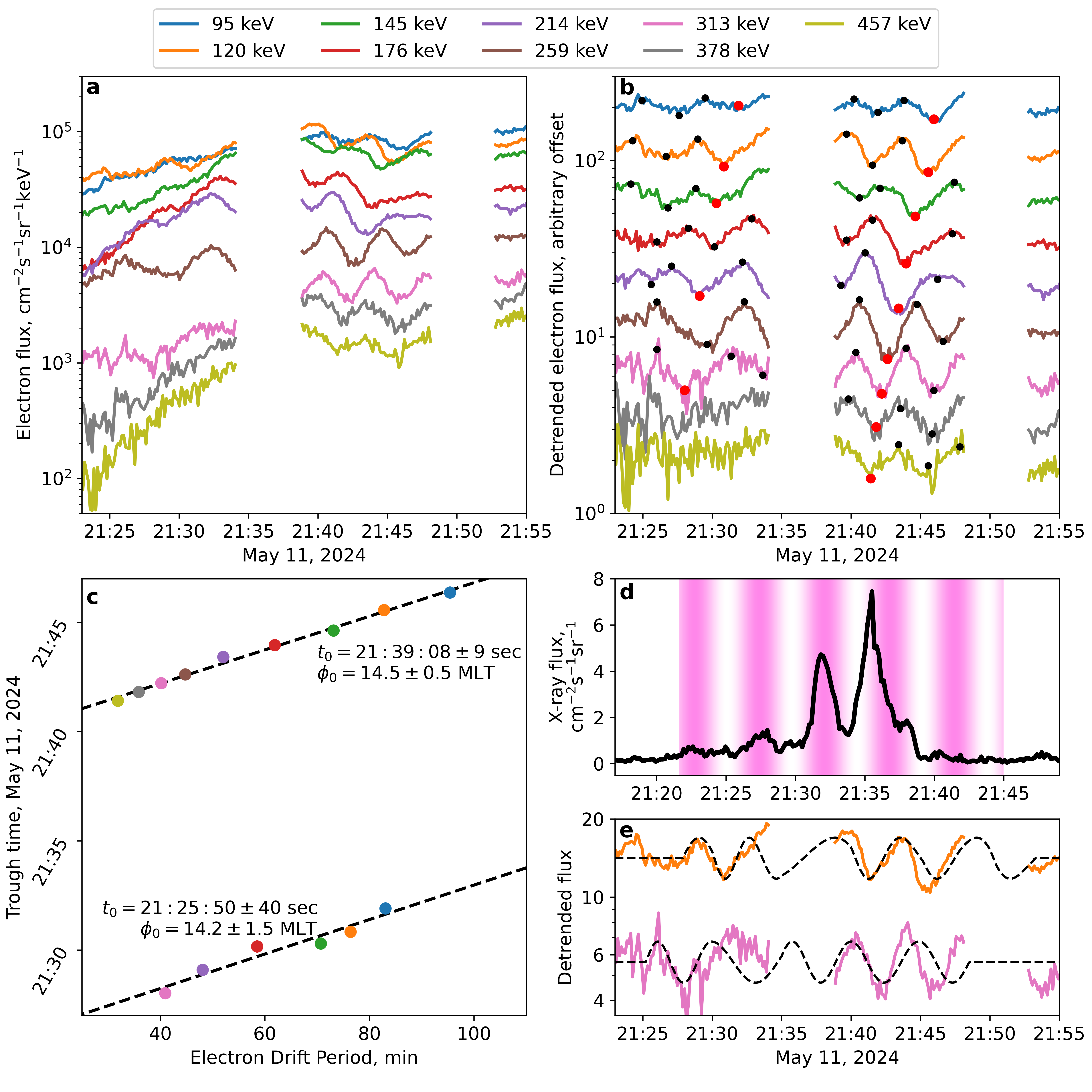}}
            \caption{Energy-dispersed electron flux modulations observed by Arase and comparison with balloon X-ray data. (a) Electron flux time series measured by Arase in multiple energy channels (95-457 keV), showing modulation during the interval of interest. (b) Detrended electron flux in the same energy channels with identified peaks and troughs (black and red dots). (c) Times of troughs identified with red dots in panel (b) plotted as a function of corresponding electron drift periods, with linear fits used to determine the modulation origin in both time and magnetic local time using equation (1). (d) Balloon-borne X-ray flux with shaded bands indicating modeled flux modulation intervals derived from the drift dispersion fit. (e) Detrended electron flux at two representative energies (120 keV in orange, 313 keV in pink) compared to the modeled modulation pattern traced to the Arase satellite location (black dashed line). }
            \label{fig:Arase_fit}
        \end{figure}

        Based on these observations, it is possible to link the ULF wave activity to indirectly modulating electron precipitation. In this event, we hypothesize that ULF waves located in the post-noon sector created patches of drift-phase bunched electron fluxes that cross the K-P limit (Figure~\ref{fig:Arase}d). This locally modulates chorus wave growth and causes periodic chorus-driven energetic electron precipitation. Note, however, that the Arase satellite measurements occurred approximately two hours later in MLT than the balloon observations (Figure~S1). Consequently, the energy-dispersed patches of drift-phase bunched energetic electrons observed by Arase must have originated as a result of ULF wave-particle interactions at earlier local times. Figure~\ref{fig:Arase_fit} presents an analysis of the energy-dependent drift dispersion observed at Arase to determine the time and MLT location of the ULF wave-particle interaction that created the flux modulations.

        The top row of Figure~\ref{fig:Arase_fit} shows the energetic electron flux measurements from Arase across multiple energy channels.  Figure~\ref{fig:Arase_fit}a displays the electron flux time series. Figure~\ref{fig:Arase_fit}b shows the corresponding detrended electron flux obtained by removing a 20-minute rolling average from the electron flux data from Figure~\ref{fig:Arase_fit}a. We use detrended flux to more easily identify the timing of the flux peaks and troughs needed for tracing their origin. Notably, all energy channels between 95~keV and 378~keV exhibit clear, well-defined modulations accompanied by pronounced energy dispersion, with higher energies arriving at the satellite earlier and having shorter oscillation periods (in the reference frame of the spacecraft). To quantify the source location of these modulations, we focus on two clear flux troughs marked by overlaid red scatter points in Figure~\ref{fig:Arase_fit}b. Note that two energy channels were excluded from the fitting dataset of the earlier trough, as they fall outside of the 3-sigma range with the fit line. These troughs, situated near the beginning and end of the modulation interval, are fitted using equation~(1) to estimate the time and MLT location where the modulation was initiated. Figure~\ref{fig:Arase_fit}c shows these fits with the identified $t_0$ and $\phi_0$ parameters for both troughs. 
        
        Figure~\ref{fig:Arase_fit}c shows that equation (1) provides a good fit to the data. Importantly, both fits yield consistent values for the MLT of the ULF modulation origin within 1-$\sigma$ uncertainty: $\phi_0=$14.5$\pm$0.5 MLT and $\phi_0=$14.2$\pm$1.5 MLT. This identified origin point in MLT is also consistent with the location of the balloon when it observed the modulated precipitation ($\sim$13.5~MLT, cf. Figure~S1). This consistency suggests that the observed flux modulation---and by extension, the modulation of electron precipitation---is primarily a temporal phenomenon generated by ULF wave-particle interaction as injected electrons drift into a narrow region of enhanced ULF wave activity near noon. There, the fluxes are increased or lowered as a result of drift-phase bunching relative to the phase of the wave. These patches of electron flux then continue to drift freely, producing an energy-dispersed flux modulation when measured by Arase. 

        The fits arising from the model allow us to verify the source of the electron flux modulation at its origin. Here, we modeled the resulting flux perturbation as a cosine wave with troughs at 21:25:50 UT and 21:39:08 UT, following the fitting results. Figure~\ref{fig:Arase_fit}b shows that these are not adjacent troughs; two additional troughs occur between them, not very clearly visible in the Arase data due to a data gap. The resulting flux perturbation can be described by the functional form $\cos(1.34 t + 0.910)$, where $t$ is the time in minutes since 21:00 UT on May 11, 2024. Figure~\ref{fig:Arase_fit}d overlays this model electron flux perturbation as periodic pink bands on the balloon X-ray flux, with solid pink indicating flux peaks and white indicating troughs. Notably, all four X-ray peaks, which correspond to enhanced precipitation, align with peaks in the modeled electron flux, consistent with the interpretation that increased fluxes near or above the K-P limit local to the balloon promote increased chorus wave growth and thus enhanced electron precipitation.

        This model was then propagated to Arase. Figure~\ref{fig:Arase_fit}e compares the modeled flux perturbation with the observed Arase electron fluxes at two representative energies, showing good agreement despite the model's simplicity. A slight timing discrepancy between the model and Arase observations is evident for the last peak at both Arase and the balloon, with the data peaks occurring approximately one minute earlier than the model prediction. We attribute this mismatch to the simplistic nature of the cosine model, as the ULF wave-particle interactions may produce non-uniformly spaced particle flux enhancements depending on the exact characteristics of the ULF wave form.
       
        
    \section{Discussion}
        \begin{figure}
            \centering
            \makebox[\textwidth][c]{\includegraphics[width=1.2\linewidth]{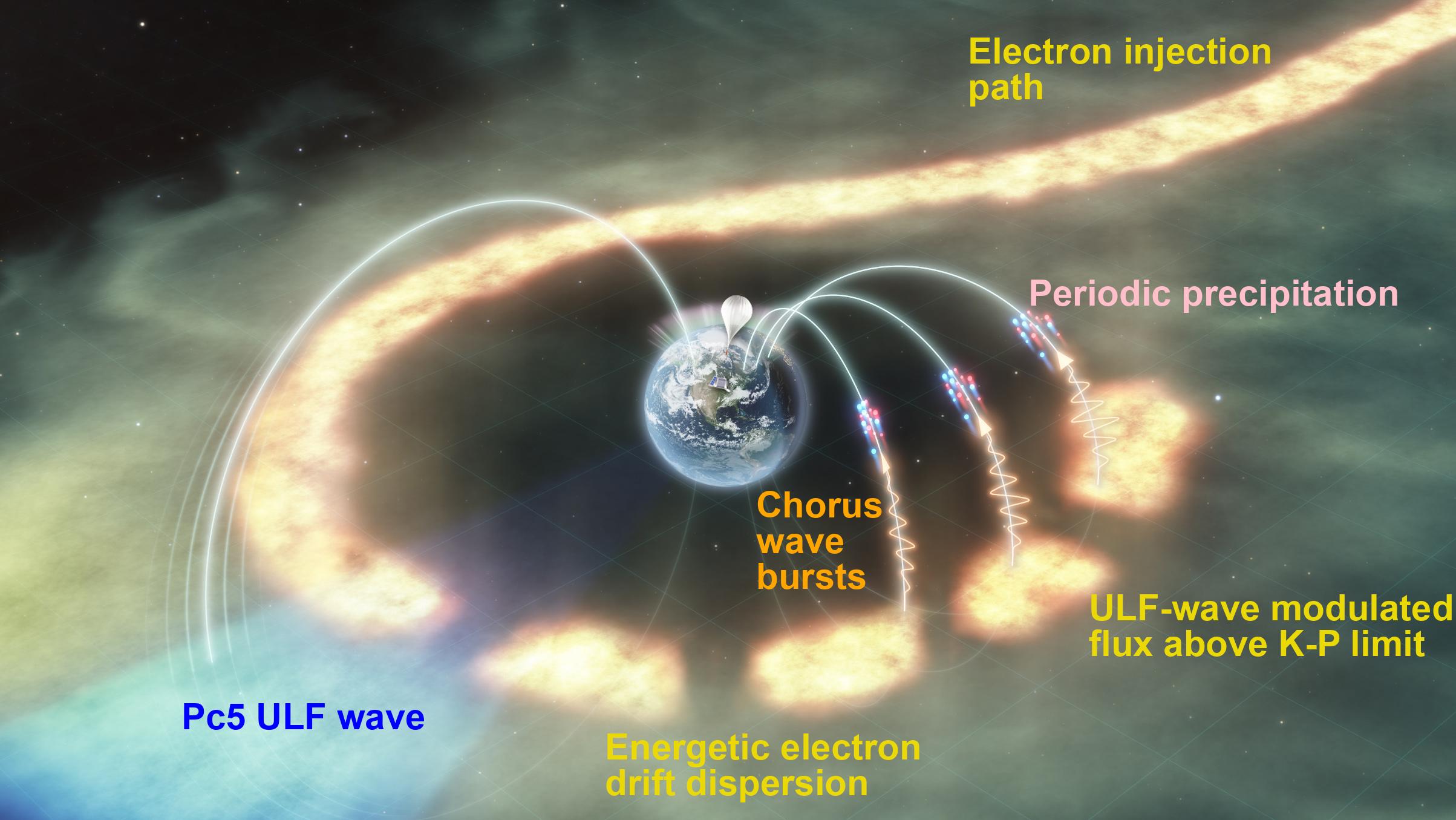}}
            \caption{Schematic illustration of ULF wave-modulated electron fluxes and resulting modulated energetic electron precipitation. Energetic electrons injected into the outer radiation belt undergo periodic modulation by ultra-low-frequency (ULF) waves-particle interaction in the post-noon sector, creating drift-phase bunched regions of enhanced electron flux that drift eastward around the Earth. These modulated electron flux populations serve as a periodic seed source for chorus wave growth near the Kennel-Petschek limit, which leads to enhanced pitch-angle scattering and modulated electron precipitation into the upper atmosphere. }
            \label{fig:Schematic}
        \end{figure} 
        
        Our results support a model in which ULF waves indirectly drive modulated electron precipitation by periodically modulating energetic electron fluxes near the Kennel-Petschek (K-P) limit as a result of drift-phase bunching. Figure~\ref{fig:Schematic} illustrates the progression of the ULF wave modulation of the drifting energetic electrons, the excitation of enhanced chorus waves at the K-P limit, and the resulting precipitation. Initially, energetic electrons transported from the plasmasheet enter the inner magnetosphere, as evidenced by the enhanced $>$80~keV flux observed near 21:20~UT in Figure~\ref{fig:Arase}b. This injection increases the background energetic electron fluxes sufficiently to approach the K-P limit. Active within a localized post-noon MLT sector, ULF waves periodically modulate the newly injected electron flux, creating quasi-periodic drift-phase bunched patches of enhanced energetic electron flux (Figures~\ref{fig:Arase}b). Within these patches, fluxes surpass the K-P limit (Figures~\ref{fig:Arase}d,e), thereby significantly amplifying the local chorus wave growth rate and local chorus wave power (Figure~\ref{fig:Arase}c). Enhanced chorus activity increases electron pitch-angle scattering, driving intense and quasi-periodic electron precipitation bursts into the atmosphere (Figure~\ref{fig:Arase}a). The resulting periodic precipitation pulses originate at or near the ULF wave-particle interaction region, where the balloon measurements provide fortuitous in-situ observations directly beneath the active ULF wave-particle interactions. Finally, these electron flux structures then drift eastward along the drift trajectory, energy dispersing over time and later being measured in Arase satellite data with an approximate 6~min delay from their detection at the balloon. 

        Interestingly, \citeA{Olifer_2023} studied somewhat similar transient increases of the energetic electron flux above the K-P flux limit. In the flux statistics shown in their letter, \citeA{Olifer_2023} showed that the electron flux at energies $\sim$65~keV can often exceed the K-P limit across a range of local times from midnight to noon along their drift trajectory, causing associated electron precipitation. Thus, even modest flux variations, like those observed here and induced by ULF wave-particle interaction and drift-phase bunching, can exceed the K-P threshold for a sufficient amount of time to produce a sequence of enhanced electron precipitation.


    \section{Conclusions}
        In this letter, we showed how ultra low frequency (ULF) waves can indirectly modulate storm-time energetic electron precipitation by periodically modulating electron fluxes of the newly injected $\sim$80-300~keV population near the Kennel-Petschek (K-P) limit. Intervals where the local electron flux exceeds the K-P limit cause enhanced chorus wave growth and produce quasi-periodic atmospheric precipitation bursts. Overall, the findings of our paper can be summarized as follows: 
        
        \begin{enumerate}
            \item Coordinated balloon-borne Timepix X-ray and ground-based riometer measurements revealed four $\sim$4~min period modulation of electron precipitation that were phase-coincident and conjugate to monochromatic Pc5 ULF oscillations during the recovery phase of the May 2024 superstorm.
            \item Concurrent Arase energetic electron flux data showed energy-dispersed ($\sim$80-300~keV) flux modulations with the same $\sim$4~min periodicity; higher-energy electrons in each modulation were observed earlier at Arase, consistent with drift dispersion from a localized modulation region.
            \item Linear fits to the energy-dependent timing of the dispersed election flux patches identified the modulation origin in the post-noon sector ($\sim$14 MLT), collocated (within uncertainty) with the balloon footprint. Drift trajectory modeling reproduced the observed $\sim$6~min drift delay between the balloon-observed electron precipitation peaks and the periodic Arase flux enhancements.
            \item Energetic electron fluxes at representative energies were shown to periodically exceed the K-P limit; modest ULF wave-driven flux perturbations, most likely as a result of drift-phase bunching, periodically pushed the system across the K-P threshold, enabling strong local chorus wave growth and enhanced pitch-angle scattering into the atmosphere.
            \item A periodic model for the assumed flux modulation produced by the ULF wave-particle interaction and initialized at the inferred modulation time and location reproduced the phase and spacing of the balloon X-ray peaks and the Arase flux structure, supporting the hypothesis that the ULF wave was the indirect driver of the observed precipitation modulation.
        \end{enumerate} 

        \nocite{Allanson_2019}
        
	\acknowledgments
    LO is supported by a Banting Fellowship. DZ is supported by University of Alberta SESA program. MP and MKH are supported by AFOSR GRANT FA9550-23-1-0629. IRM, COH, and GRS is supported by a Discovery grant from Canadian NSERC. We thank Dr. Kevin Pham from the NCAR High Altitude Observatory for providing MAGE simulation fields and NCAR CISL for computational resources. CARISMA is operated by the University of Alberta, funded by the CSA as a part of Space Environment Canada. Funding for operation of the NORSTAR riometers is provided by the CSA. 
	

		
	\bibliography{biblist.bib}

\end{document}